**Spin phonon interactions and magneto-thermal transport behavior in p-Si**

Paul C Lou[1], Laura de Sousa Oliveira[1], Chi Tang[2], Alex Greaney[1,3], and Sandeep Kumar[1,3,*]

[1] Department of mechanical engineering, University of California, Riverside, CA

[2] Department of physics and astronomy, University of California, Riverside, CA

[3] Material science and engineering program, University of California, Riverside, CA

**Abstract**

The spin-phonon interaction is the dominant process for spin relaxation in Si, and as thermal transport in Si is dominated by phonons, one would expect spin polarization to influence Si's thermal conductivity. Here we report the experimental evidence of just such a coupling. We have performed concurrent measurements of spin, charge, and phonon transport in p-doped Si across a wide range of temperatures. In an experimental system of a free-standing 2 μm p-Si beam coated on one side with a thin (25 nm) ferromagnetic spin injection layer, we use the self-heating 3ω method to measure changes in electrical and thermal conductivity under the influence of a magnetic field. These magneto-thermal transport measurements reveal signatures in the variation of electrical and thermal transport that are consistent with spin-phonon interaction. Raman spectroscopy measurements and first principle's calculations support that these variations are due to spin-phonon interaction. Spin polarization leads to softening of phonon modes, a reduction in the group velocity of acoustic modes, and a subsequent decrease in thermal conductivity at room temperature. Moreover, magneto-thermal transport measurements as a function of temperature indicate a change in the spin-phonon relaxation behavior low temperature.

**Keywords-** Silicon, spin-phonon interactions, thermal conductivity, magneto-thermal transport

**Introduction**

Spintronics devices — electronic devices that use spin current rather than just charge to process information — are believed to be the future of modern semiconductor devices, enabling information processing in a four-state paradigm rather than the on/off binary paradigm of current digital electronics. However, to perform this, devices must possess the ability to generate, detect, and manipulate a spin current, and so over the last decade there has been an extensive research effort devoted to advancing the fundamental understanding of spin transport that is necessary for achieving these capabilities. Much of this effort has focused on semiconductor materials with the aim of building on our exiting understanding of how to engineer these systems to exert exquisite control on their charge transport[1, 2]. Among the semiconductor materials, silicon is prime: it is first in its ubiquity of use, the depth to which it has been studied, and precision with which it can be processed — the element has become a synecdoche for the entire field of information technology. Si is also emerging as the most promising candidate for spintronic devices. It has relatively weak spin-orbit coupling, crystal inversion symmetry and zero nuclear spin, which prohibit spin relaxation by the Dyakonov-Perel mechanism, and hyperfine interactions [3]. This has stimulated research towards realizing silicon spintronic devices that has yielded tremendous progress [2]. Methods have been demonstrated for spin current injection into silicon by tunneling from a ferromagnetic electrode across a thin insulator[4, 5]. Spin can now be injected at room temperature spin injection into both n-type and p-type Si [6]. Zhang et al. [7] demonstrated all-electrical spin injection, transport, and detection in heavily n-type silicon using ferromagnetic Co/Al$_2$O$_3$ tunnel barrier contacts and high spin injection efficiency (up to ~30%) and long spin diffusion lengths (up to ~6 μm) have been achieved.

The primary spin relaxation mechanism in silicon at room temperature is electron-phonon scattering (the Elliot-Yafet mechanism). This makes Si a model material to study spin-phonon interactions. We hypothesize that by introducing spin polarization into Si will influence thermal transport, and that measuring changes in thermal conductivity could be a new route for detecting spin currents. This manuscript reports on a set of experiments and measurements that support this hypothesis.

**Experimental setup**

We devised an experimental setup with device that utilizes spin-Seebeck tunneling to inject a spin current into a 2 μm thick free standing p-doped Si beam[8]. The beam is thinner than the reported spin diffusion length of up to 6 μm [2, 6, 7].

The thermal conductivity along the axis of the beam was measured using the self-heating 3ω method [9-11] in which a sinusoidal electrical current is passed along the beam and used to both heat the beam and measure its temperature as shown in Figure 1 (a) [12, 13]. As the beam is free standing the majority of the Joule heating of the beam must be dissipated out through its ends crating a parabolic temperature distribution. There are two timescales in this system: the period of the driven current, and the thermal relaxation time. The latter of these causes a third harmonic component (frequency 3ω) in the driving voltage the magnitude of which can be used to infer the thermal conductivity. The technique relies on the solution of the one-dimensional heat conduction equation for the specimen, which is given by,

$$\rho C_p \frac{\partial \theta(x,t)}{\partial t} = \kappa \frac{\partial^2 \theta(x,t)}{\partial x^2} + \frac{I_o^2 \sin^2 \omega t}{LS}(R_o + R'\theta(x,t)), \qquad (1)$$

where $L$ and $S$ are the length between the voltage contacts and the cross-sectional area of the specimen, respectively. $\rho$, $Cp$ and $\kappa$ are the density, specific heat and thermal conductivity in the material. $R_0$ is the initial electrical resistance of the specimen at temperature $T_o$. $R'$ is the temperature derivative of the resistance $R' = \left(\frac{dR}{dT}\right)_{T_o}$ at $T_o$. $\theta(x,t) = T(x,t) - T_o$ is the temporal ($t$) and spatial ($x$) dependent temperature change, as measured along the length of the specimen, which coincides with the heat flow direction as shown in Figure 1 (a). The thermal conductivity can be expressed in terms of the third harmonic voltage $V_{3\omega}$ in the low frequency limit [9] by

$$\kappa \approx \frac{4I^3 R_o R' L}{\pi^4 V_{3\omega} S} \qquad (2)$$

when the pre-requisite condition $\frac{I_o^2 R' L}{n^2 \pi^2 \kappa S} \ll 1$ is satisfied. Sources of error include goodness of fit for $R'$ and heat loss. Heat loss is addressed by performing experiment in vacuum to eliminate convective heat loss and fabricating the device to freestanding to eliminate conductive heat loss to substrate. In our experiment the beam is coated on its top surface with a 25 nm thick ferromagnetic metallic layer. The electrical conductivity of this layer is much larger than that of the Si below it, but the reverse is true for the thermal conductivity. This means that the under the sinusoidal current most of the heat is evolved in the metallic layer, but it is conducted away in the Si beam. The resulting heat current from the magnetic film to the silicon beam carries with it a spin current.

The experimental system was prepared by using standard nanofabrication methods to create a freestanding multilayer specimen ($l$-170 μm, $w$-9 μm and $t$-2 μm) consisting of Pd (1 nm)/ Ni$_{80}$Fe$_{20}$ (25 nm)/MgO (1 nm)/p-Si (2 μm) as shown in Figure 1 (a). To fabricate the experimental setup, we took a commercially available SOI wafer with a B-doped 2 μm thick device layer having resistivity of 0.001-0.005 Ω cm. Using UV photolithography and deep reactive ion etching (DRIE), we etched the handle layer underneath the specimen region, and then patterned and etched the device layer on the front side. The silicon structure was made freestanding using hydrofluoric acid vapor etching. In the next step, surface oxide was removed by Ar milling for 15 minutes after which a 1 nm thick layer of MgO was deposited using RF sputtering. This was followed by deposition of the 25 nm thick layer of $Ni_{80}Fe_{20}$ and finally the 1nm Pd coating using e-beam evaporation. The material deposition using evaporation leads to line of sight thin film deposition. The fabricated device is shown in Figure 1 (a). The freestanding specimen overcomes the thermal conduction losses crucial for thin film thermal characterization. The MgO layer is required for efficient spin tunneling, and the Pd layer inhibits the oxidation of the $Ni_{80}Fe_{20}$ (ferromagnetic- FM) layer. The total electrical resistance of the 2 μm Si beam was comparable with the resistance of the 25 nm metal film (estimated at ~390 Ω and ~290 Ω respectively). On the other hand a ~25 nm thick layer of $Ni_{80}Fe_{20}$ has a thermal conductivity of ~20 W/m·K [14], much lower than the ~80 W/m·K of Si [15]. These relative values of electrical and thermal conductivity result is spin injection by spin-Seebeck tunneling, which has been experimentally demonstrated for Si [8]. While spin-Seebeck tunneling is assumed to be the primary mechanism of spin-polarization in the experimental system, there may also be a contribution from the spin-Hall effect (SHE) due to spin absorption when in proximity to the ferromagnet. The p-Si has been reported to exhibit the SHE[16, 17] [18-22] and inverse spin-Hall effect (ISHE) has been observed in p-Si[23]. The Rashba-Edelstein effect (REE) due to interfacial Rashba spin-orbit coupling (SOC) is the third mechanism that may lead to spin polarization in p-Si[24-27], which has also been reported in p-Si[18, 27]. These mechanisms may give rise to spin polarization this study. But, the goal of this study to elucidate the spin-phonon interactions. The changes in electro-thermal transport behavior due to spin polarization should be observable in resistance and third harmonic voltage measurements.

(Figure 1)

**Results and discussion**

Measurements of transport properties under varying temperature and magnetic field were made using a Quantum Design physical property measurement system (PPMS). In order to establish and calibrate a reliable measurement regime, first, the current and frequency dependence of the $V_{3\omega}$ signal was measured at room temperature. The former is plotted in in Figure 1 (b) and shows a cubic relationship, while the latter is plotted in Figure 1 (c) for frequencies from 2–20 Hz. This verifies that $V_{3\omega}$ is frequency independent which is an important requirement for the $3\omega$ method [9]. To calculate the thermal conductivity, one must also know $R'$. This was determined by taking the gradient of a polynomial fit to the resistance at zero magnetic field measured at temperatures between 50 and 310 K, as shown in Figure 1 (d). For all subsequent measurements, those taken using a sinusoidal current were performed with the RMS amplitude of 1 mA, and direct current measurements were performed at 10 µA.

To test the dependence of transport properties on applied magnetic field, both $V_{3\omega}$ and DC resistance were measured while sweeping the magnetic field from -3 T to 3 T. The DC resistances exhibits negative magnetoresistance (MR) of -1.11% for a magnetic field of 3T as shown in Figure 2 (a). Since both $Ni_{80}Fe_{20}$ and p-Si layers contribute to MR, we also measured MR of a control specimen, a Si beam with no magnetic layer, which showed positive MR (Supplementary Figure S1). We used the $V_{3\omega}$ to determine the zero-field thermal conductivity, $\kappa$ of the bilayer to be ~59.21 W/m·K, a value that agrees with the thermal conductivity values for highly doped Si found in the literature[15, 28]. The magnetic field and temperature dependence of $R'$ are shown in Figure 2 (c) with applied field of zero and 1.25 T (corresponding to the knee in MR). We estimate that $R'$ increases by ~4% for 1.25 T as compared to zero field. The fluctuations are due to instrumental noise and curve fitting. We decided to neglect the change in $R'$ due to magnetic field for thermal conductivity calculations since it does not change the trend observed in the thermal conductivity measurements. This simplification underestimates the calculated magneto-thermal conductivity, since the $R'$ increases as a function of magnetic field.

(Figure 2)

To elucidate the magneto-thermal transport behavior, we performed electrical and thermal transport measurements while sweeping an applied out-of-plane magnetic field at temperatures 300 K, 200 K, 150 K, 100 K and 50 K. The results in Figure 2 (a) show a negative MR in $Ni_{80}Fe_{20}$/p-Si bilayer for all the temperatures. The negative MR is calculated to be -1.76% at 200 K and changes to its most negative value of -2.52% at 50 K as shown in Figure 2 (a). At all the temperatures, the knee in MR at ~1.25 T corresponds to the saturation magnetization of the $Ni_{80}Fe_{20}$ layer as stated earlier. The negative MR after the saturation magnetic field is attributed to the reduction in magnon population at higher magnetic fields and lower temperatures[29]. The Figure 2 (b) shows the dependence of $\kappa$ with applied magnetic field at a number of different temperatures similar to MR measurement. In all cases $\kappa$ decrease with both a positive and negative applied magnetic field. The $\kappa$ decreases rapidly with introduction of out-of-plane magnetic field and reaches a minimum at 1.25 T, which is the out-of-plane magnetic saturation of $Ni_{80}Fe_{20}$. While $\kappa$ is reduced for both field directions, it shows some slight asymmetry with magnetic field and is lower at negative direction, showing saturation-dependent anisotropic thermal conductivity similar to that of MR. The largest $\kappa$ was observed at 150 K under no field as well as maximum change of 3.39 % under applied magnetic field. At 100 K, magneto thermal transport behavior experiences a minimal change (1.81 %), which is also supported by the magnetic field dependent $V_{3\omega}$ response (supplementary Figure S2). The raw $V_{3\omega}$ response as a function of magnetic field for the lowest four temperatures is shown in Supplementary Figure S2, and corresponding magnetic field dependent $\kappa$ in Supplementary Figure S3.

The observed asymmetric magneto-thermal transport behavior can arise from the phonon anharmonicity[30, 31] due to phonon induced diamagnetic force in the presence of magnetic field. In addition, optical absorption studies reported on boron doped silicon found that Zeeman splitting of the excited states of heavy hole $\left(0.5\ m_0, j_z = \pm\frac{3}{2}\right)$ is asymmetric and anisotropic[32]. But these mechanisms do not explain the absence of asymmetry at and below 200 K. Hence, we propose that spin-phonon interactions are the underlying reason for the asymmetric behavior and asymmetry disappears due to change in spin relaxation behavior at lower temperatures. The resistance and thermal transport measurements suggest different effects on charge and thermal transport due to spin polarization. This can be rationalized by remembering that the electrical transport in the specimen has contributions from both $Ni_{80}Fe_{20}$ and p-Si layers while thermal transport is dominated by p-Si. Since, the thermal transport in Si is phonon-mediated even for heavily doped Si used in these experiments [15] the observed magneto-thermal transport behavior must be due to interaction between phonons and spin, or the spin polarization of the p-Si layer altering the Si's phonon properties

To verify that the magneto thermal transport behavior is not an artifact of the $Ni_{80}Fe_{20}$ layer, or the interface between the Pd/$Ni_{80}Fe_{20}$ layers, the transport behavior was measured for a second control specimen consisting of a Pd/$Ni_{80}Fe_{20}$ thin film supported on a Si oxide membrane. The results are plotted in Supplementary Figure S4, and show the negative MR of ~0.54% — considerably smaller than that of the Pd/Ni$_{80}$Fe$_{20}$/MgO/p-Si multilayer specimen. This agrees with a recent report on Ni$_{80}$Fe$_{20}$/p-Si bilayer[19]. Furthermore, this control specimen displayed a asymmetric reduction in V$_{3\omega}$ as a function of magnetic field. This indicates an increase in thermal conductivity, which consistent with the change in resistance if heat conduction is dominated by electrons. We carried out similar experiments on the p-Si control specimen with no magnetic layers as shown in Supplementary Figure S1. Note that in the control specimen, V$_{3\omega}$ of p-Si alone is not affected by magnetic field. For in-plane thermal conduction, we estimated that the 2 μm p-Si is 6000 times $\left(R_{thermal} = \frac{l}{\kappa A}\right)$ more thermally conducting than the 25 nm Ni$_{80}$Fe$_{20}$ layer. Together these control experiments clearly show that the observed behavior presented in case of Pd/Ni$_{80}$Fe$_{20}$/MgO/p-Si multilayer specimen is not exclusive to the p-Si and Ni$_{80}$Fe$_{20}$ layers alone, nor from effects at the interface between Pd and Ni$_{80}$Fe$_{20}$.

The spin-phonon interaction results in creation or annihilation of a phonon to accept/provide the energy associated with a spin flip and can remove or scatter a band of phonons from the heat current[33] and reduce the thermal conductivity similar to the behavior observed in these experiments. Raman spectroscopy can provide additional insights in the spin-phonon relaxation behavior and support the magneto thermal transport measurements reported earlier. We acquired the Raman spectrum of Si on the same set of devices used for thermal transport measurements as shown in Figure 2 (d). We modified the FM layer to 10 nm of Ni$_{80}$Fe$_{20}$ to enhance the Si Raman signal. The Raman laser was also used to create a temperature gradient across the specimen for spin-Seebeck tunneling. For a 300 μm Si substrate the Γ-optical phonon peak appears at wavenumber of 521 cm$^{-1}$ for a laser power of 60 mW. In the smaller 2 μm p-Si without any FM layer, the Γ-phonon peak is also observed at 521 cm$^{-1}$ for a laser power of 5 mW, but shifts to 519 cm$^{-1}$ under laser power of 60 mW due to heating from laser. For Pd (1 nm) / Ni$_{80}$Fe$_{20}$ (10 nm) /

MgO (1 nm) / p-Si (2 µm), we observe a red shift from 521 cm$^{-1}$ to 516 cm$^{-1}$ in the Raman spectra for Γ-optical phonons. The laser power will heat Ni$_{80}$Fe$_{20}$ and cause spin-Seebeck tunneling to the p-Si layer underneath. The red shift in optical phonon frequency supports our hypothesis of spin polarization and spin-phonon interactions mediated thermal transport behavior. The Raman spectroscopy measurement indicate phonon softening due to spin polarization. These experimental observations can be further supported by the simulations.

To ascertain the influence of a net spin concentration on Si's phonon properties, a sweep of density functional theory (DFT) calculations were performed to compute phonon dispersion and group velocities in the presence of an increasing imbalance of spin-up vs spin-down electrons. These calculations were performed by finding the ground state electron density subject to the constraint of a small fixed net spin concentration (*i.e.*, a triplet state), a calculation that consistent with DFT as a ground state method. The dynamical matrix of silicon containing a net spin concentration was computed for spin concentrations from zero to 4.6×10$^{-3}$ per valence electron using the plane-wave DFT Vienna *Ab-initio* Simulation Package (VASP)[34-37]. All calculations were performed in the local density approximation (LDA)[38] using the project-augmented wave (PAW) method[39, 40], with a Monkhorst-Pack grid of 3x3x3 irreducible *k*-points and a plane wave energy cutoff of 1200 eV. The interatomic force constants (the hessian matrix) were computed from a single 3x3x3 super-cell constructed from the Si primitive cell with atomic displacements implemented by Phonopy[41] based on the structure symmetry of Si (Fd-3m). The equivalent perfect super-cell was found to be relaxed to within 1x10$^{-3}$ eV/Å ionic tolerance and a 1x10$^{-6}$ eV electronic tolerance. The phonon dispersion and phonon group velocities were also computed using Phonopy. The change in phonon velocities and phonon radiance as a function of temperature and spin concentration was determined by integrating group velocities weighted by phonon energy and occupation for all phonon branches on a 101x101x101 *k*-point grid over the entire Brillouin zone. The phonon radiance is given by:

$$I = \frac{1}{4\pi} \sum_p \frac{1}{8\pi^3} \int_{BZ} dk^3\, \omega_p \hbar\, v_p\, n,$$

where $\omega_p$ and $v_p$ are the angular frequency and group velocity (magnitude) of phonon modes with wave vector $k$ and polarization $p$, $n$ is the equilibrium occupancy of the modes given by the Bose-Einstein distribution, and the integral is performed over the first Brillouin zone and summed over all polarizations. The phonon radiance is the intrinsic energy flux in the phonon gas, the energy flux per solid angle at when the system is at equilibrium — it is a materials' capacity for ballistic transport of heat before the imposition of resistive scattering processes. The computed phonon properties are summarized in Figure 3 a-f. Forcing the spin polarization puts electrons into anti-bonding states, softening the Si-Si bonds. Consequently, the phonon band structures show a monotonically increasing red shift in the frequency of optical gamma point phonons with increasing spin polarization, a prediction that is consistent with supporting the Raman spectroscopy measurements. Accompanying the red shift of Γ-optical phonons is a reduction in the group velocity of the transverse acoustic phonons (plotted explicitly in Figure 3 d). This retardation of acoustic phonons reduces the phonon radiance and so would, by itself, reduce thermal conductivity before any additional thermally resistive processes from acoustic phonon scattering in spin relaxation processes. Figures 3-e and f shows the fractional reduction in phonon radiance (computed from Boltzmann transport theory using the single relaxation approximation) as a function of sweeping temperature and sweeping through spin concentration. The overall trends show a marked similarity with the experimentally measured magneto-thermal transport results presented in this work indicating that phonon softening can contribute to the measured results, but that there are probably also additional phonon scattering contributions due to phonon participation in spin relaxation processes. Note that these computational results only indicate a connection between the stiffness of Si and the presence of a spin imbalance, not necessarily a spin current. Using magneto thermal transport, Raman spectroscopy and simulations, we have established that spin-phonon interactions changes the thermal transport behavior.

(Figure 3)

**Conclusion**

Using magneto-thermal transport measurements, Raman spectroscopy and DFT simulations, we have demonstrated the spin-phonon interactions using magneto-thermal characterization in p-Si. The self-heating 3ω method is used for magneto-thermal characterization to uncover the spin mediated thermal transport. Spin polarization in p-Si leads to phonon softening, which we believe is one of the underlying cause of magneto-thermal transport behavior reported in this article. The spin polarization reduces when the magnetization of $Ni_{80}Fe_{20}$ layer is saturated since spin absorption is reduced. This causes the peak observed in magneto-thermal measurements. In addition, the decrease in the temperature changes the phonon occupation and spin-phonon relaxation behavior causing asymmetric magneto-thermal transport behavior to disappear at low temperatures. The self-heating 3ω method can be extended to discover the spin mediated thermal transport and spin-phonon interactions in other semiconductor and metallic thin films. In addition, the spin-phonon interactions presented in this work may allow thermal manipulation of spin current. These experimental results strengthen the foundation of silicon spintronics and may lay the ground for silicon spin-caloritronics.

**Acknowledgments**

We thank Prof. Jing Shi (UCR) and Prof. Ward Beyermann (UCR) for discussions and inputs. This work used the Extreme Science and Engineering Discovery Environment (XSEDE), which is supported by National Science Foundation grant number ACI-1053575.

**List of Figures:**

Figure 1. (a) SEM micrograph of the experimental setup with freestanding specimen, (b) cubic relationship between heating current and $V_{3\omega}$ response (c) $V_{3\omega}$ response as a function of frequency of heating current, and (d) resistance as a function of temperature showing $R'$ at 300 K, 2000 K, 150 K, 100 K and 50 K.

Figure 2. (a) Magnetoresistance behavior of the specimen as a function of temperature from 300 K to 50 K, (b) change in thermal conductivity, $\kappa$ as a function of magnetic field and temperature, (c) The $R'$ as a function of temperature between 50 K and 300 K for zero field and 1.25 T. The difference between 0 T and 1.25 T is approximately 4%, and (d) Raman spectra of bulk Si wafer, 2 μm p-Si device layer and 10 nm $Ni_{80}Fe_{20}$/ 1 nm MgO/ 2 μm p-Si specimen showing red shift due to spin polarization.

Figure 3. Plots showing DFT predicted phonon properties as a function of spin concentration. In plots (a–e) results are plotted for spin concentrations of 0, 0.046, 0.23, 0.46, 1.39, 2.3, 3.2, and 4.6 per $10^3$ valence electrons, with the plot color from blue through green to yellow going in order of increasing spin concentration. Plot (a) shows the phonon dispersion, plot (c) the phonon density of states with inset (c) the showing the reduction in the gamma point optical phonon frequency as a function of spin. Plot (d) shows the group velocity of acoustic phonons along the {100} direction for different spins from which it can be seen that there is a marked softening of the transverse acoustic modes. Plot (e) shows the resulting predicted fractional change in phonon radiance due to spin ($\kappa(T,H)/\kappa(T,0) - 1$) as a function of temperature. This was obtained from Boltzmann transport theory assuming the single relaxation approximation. Plot (f) shows the predicted variation in thermal conductivity when sweeping the spin concentration at temperatures of 300,

250, 200, 150, 100, and 50 K (colored from dark red to orange with increasing temperature. It is clear from this data that spin induced phonon softening can account for some of the observed reduction in thermal conductivity, but not all of it indicating that probably increases phonon scattering due to phonons participating in spin relaxation processes.

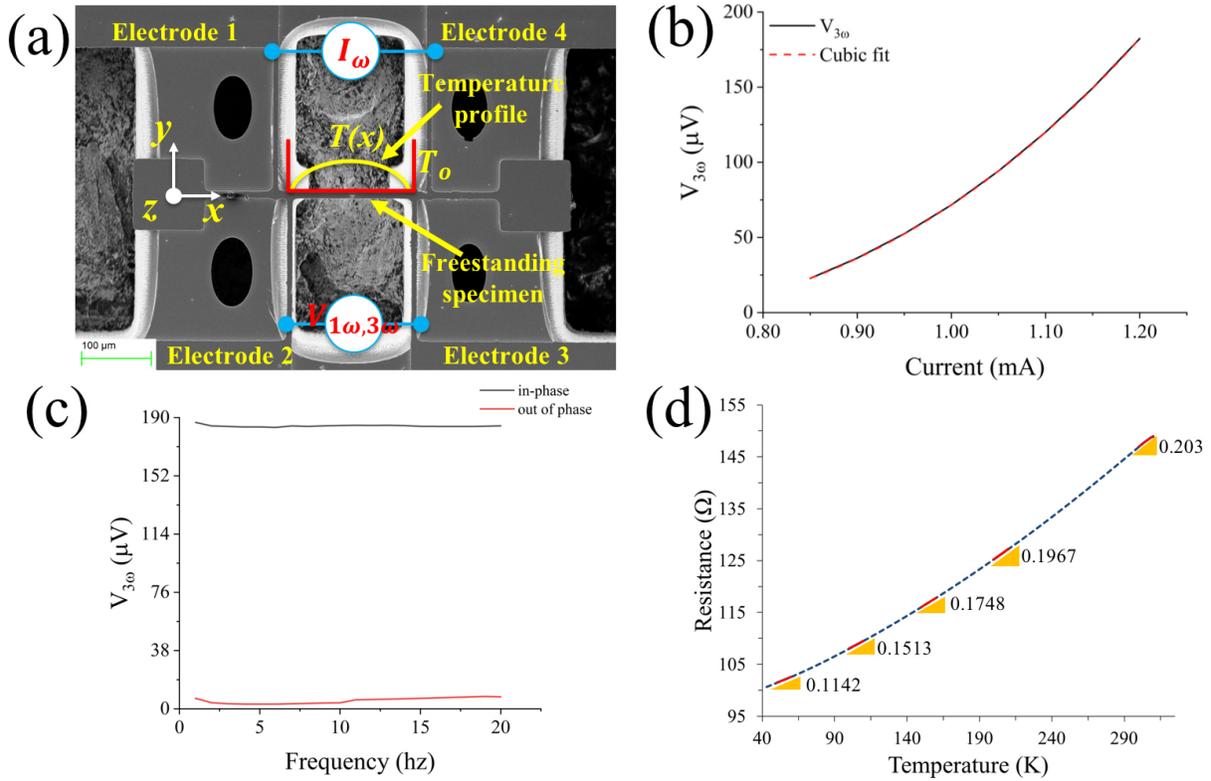

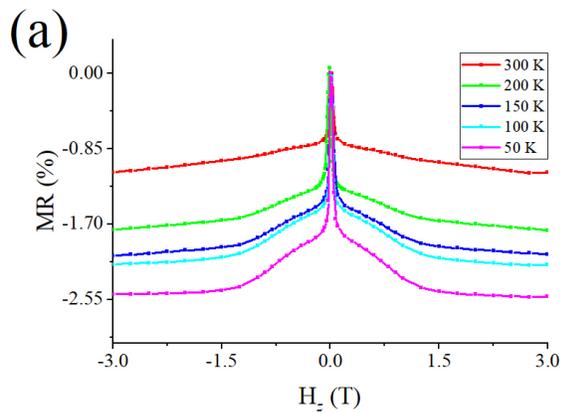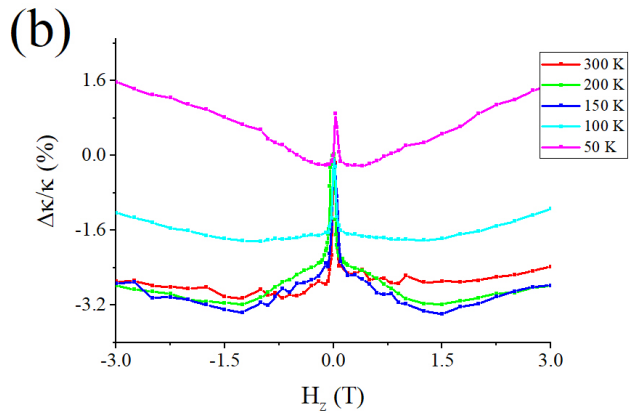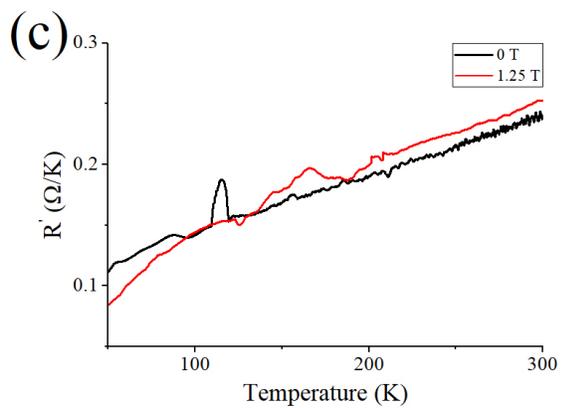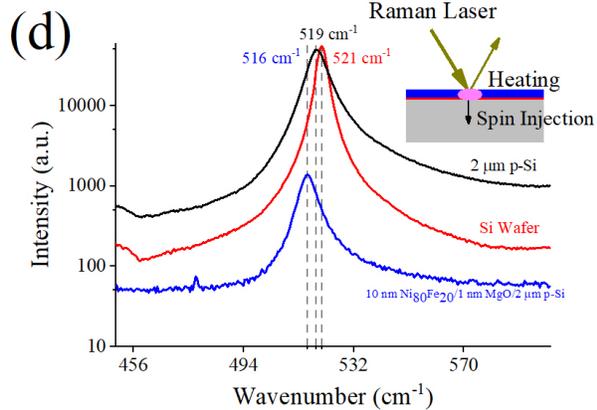

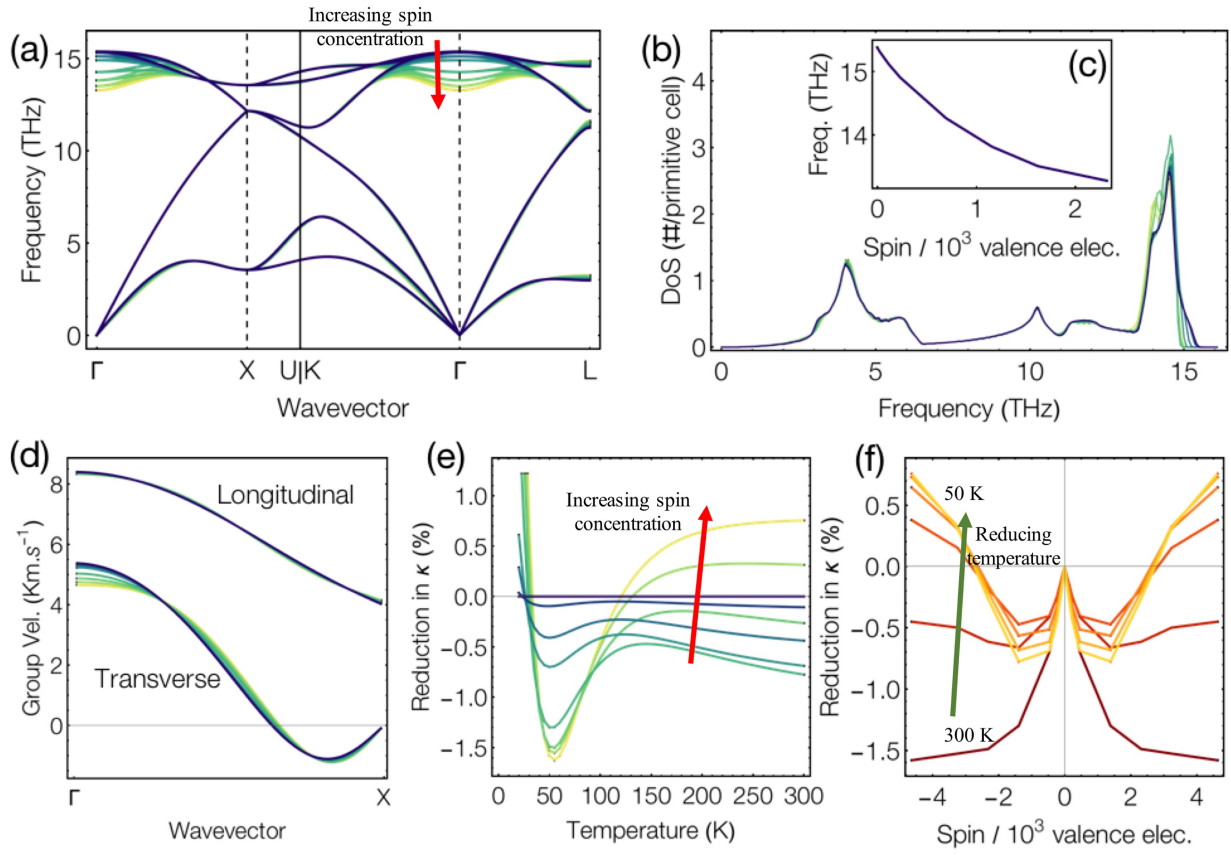

**Supplementary Figures –**

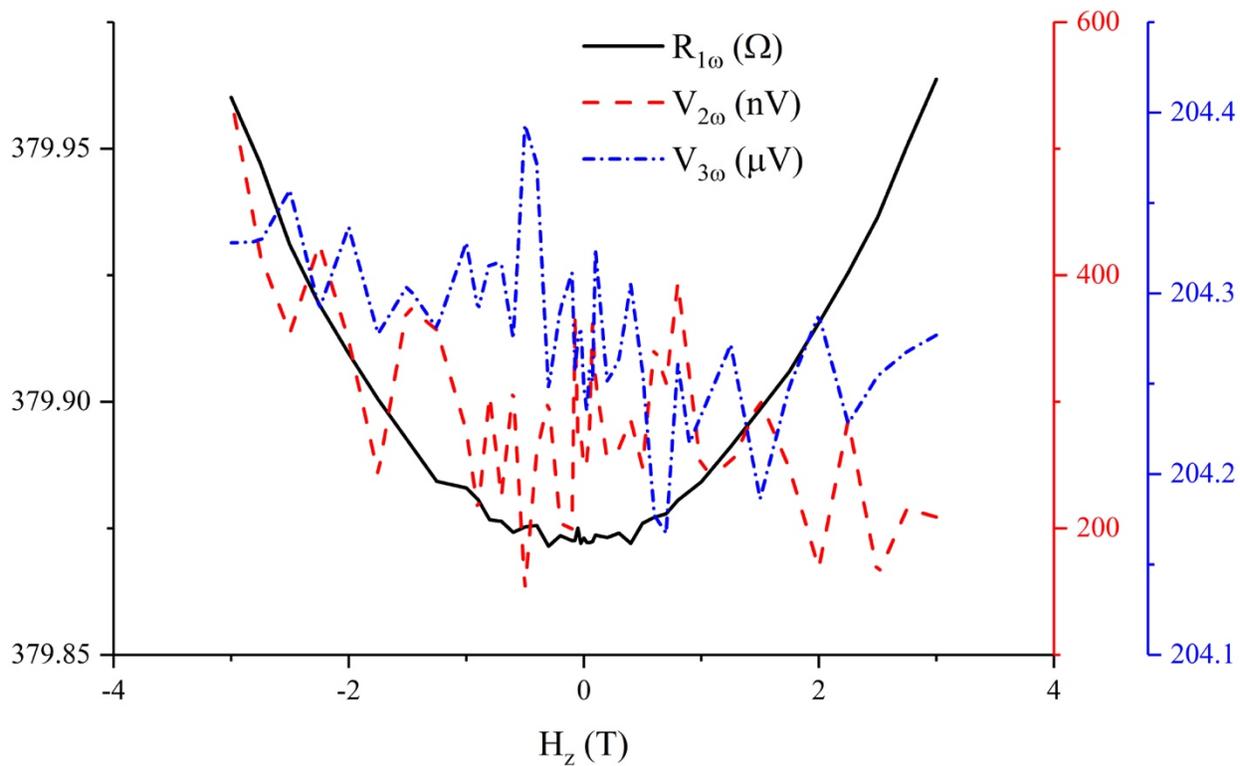

Supplementary Figure S1. Magnetic field dependent resistance, $V_{2\omega}$ response and the $V_{3\omega}$ response of a bare Si specimen at 300 K showing positive MR.

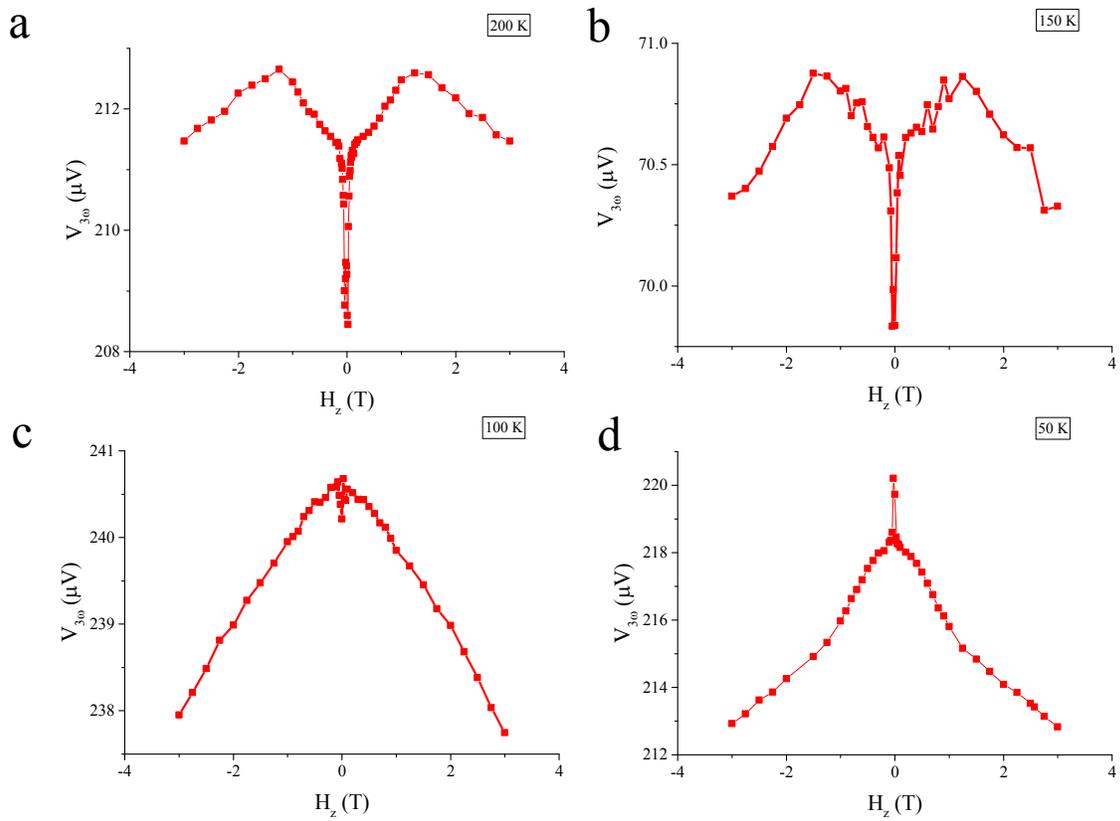

Supplementary Figure S2. The $V_{3\omega}$ as a function of magnetic field (a) 200 K, (b) 150 K, (c) 100 K and (d) 50 K. During the experiment $V_{3\omega}$ decreases as the thermal conductivity increases, so we increased the heating current to 1.5 mA to improve the signal-to-noise ratio as the temperature is reduced below 150 K.

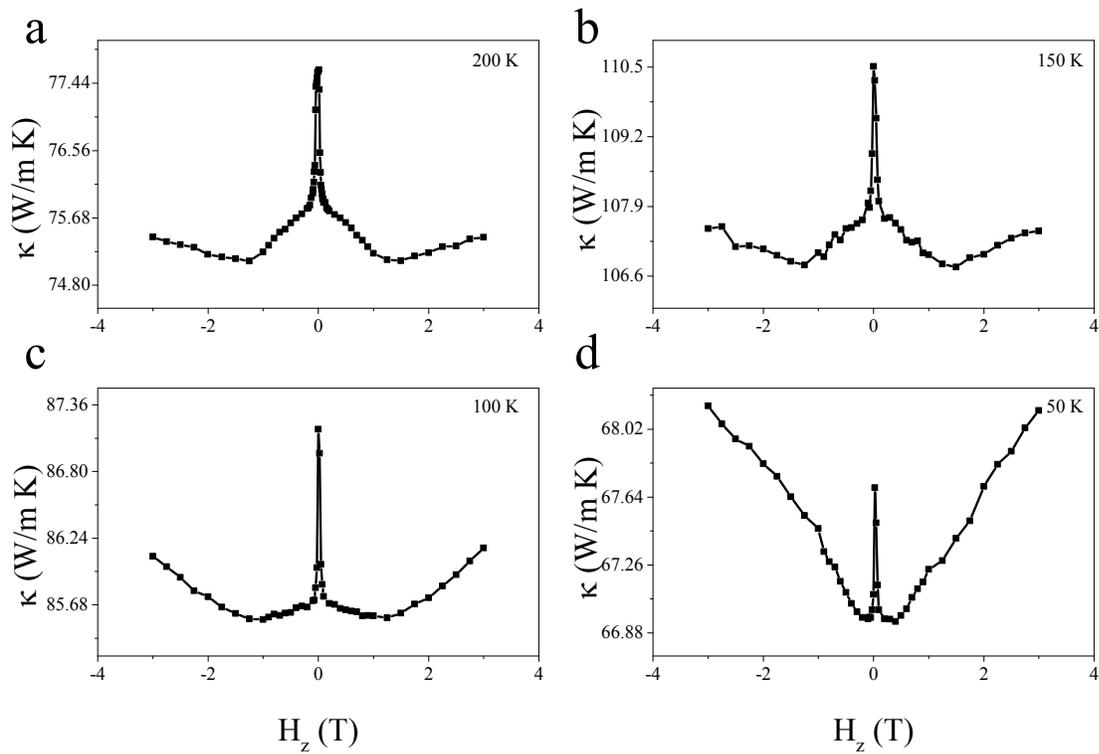

Supplementary Figure S3. The thermal conductivity as a function of magnetic field (a) 200 K, (b) 150 K, (c) 100 K and (d) 50 K.

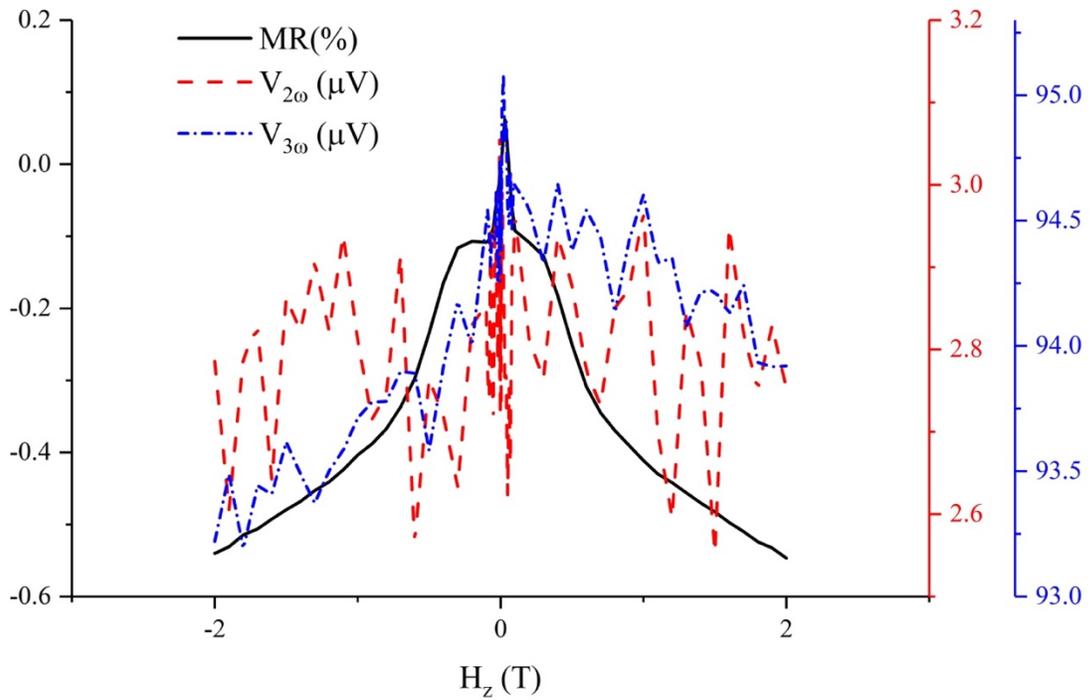

Supplementary Figure S4. Magnetic field dependent magnetoresistance (%), $V_{2\omega}$ response and the $V_{3\omega}$ response of a Pd/Ni$_{80}$Fe$_{20}$ specimen at 300 K showing negligible signal (mostly noise) and no magnetic field dependence. The $V_{3\omega}$ reduces when the magnetic field is applied, which is opposite to the Ni$_{80}$Fe$_{20}$/MgO/p-Si specimen.